\documentclass[conference]{IEEEtran}
\usepackage[margin=0.7in]{geometry}
\IEEEoverridecommandlockouts
\usepackage{cite}
\usepackage{amsmath,amssymb,amsfonts}
\usepackage{algorithmic}
\usepackage{graphicx}
\usepackage{textcomp}
\usepackage{xcolor}
\def\BibTeX{{\rm B\kern-.05em{\sc i\kern-.025em b}\kern-.08em
    T\kern-.1667em\lower.7ex\hbox{E}\kern-.125emX}}

\usepackage{amsthm}
\usepackage{amssymb}
\usepackage{amsmath}

\newtheorem{proposition}{Proposition}

\usepackage{amsmath,amsfonts,amsthm,bm} 
\usepackage{cite}
\usepackage{booktabs}
\usepackage{float}
\makeatletter

\newcommand{\Rmnum}[1]{\expandafter\@slowromancap\romannumeral #1@}
\makeatother

\DeclareMathOperator{\minimize}{minimize}

\DeclareMathOperator{\subto}{s.t.}
\usepackage{color, soul} 
\usepackage{graphicx} 
\usepackage{graphics} 
\usepackage{amsmath} 
\usepackage{verbatim}
\usepackage{mathrsfs}
\usepackage{caption}
\usepackage{multirow}
\usepackage{float}
\usepackage{epstopdf}
\renewcommand{\Re}{\operatorname{Re}}
\usepackage{array}
\DeclareMathSizes{12}{10}{8}{6}
\usepackage{caption}
\usepackage{subfigure}
\usepackage{color}
\definecolor{gray}{rgb}{0.6, 0.6, 0.6}
\definecolor{blue}{rgb}{0.0, 0.5, 0.89}
\begin{document}

\title{Near-optimal Detector for SWIPT-enabled Differential DF Relay Networks with SER Analysis\\
\thanks{This work was supported by the Hong Kong Research Grants Council under GRF project no. 16233816.}
}

\author{\IEEEauthorblockN{ Yuxin Lu and Wai Ho Mow}
\IEEEauthorblockA{\textit{Department of ECE, The Hong Kong University of Science and Technology} \\
Hong Kong SAR, China\\
Email: \{ylubg, eewhmow\}@ust.hk} 
}

\maketitle

\begin{abstract}
In this paper, we analyze the symbol error rate (SER) performance of the simultaneous wireless information and power transfer (SWIPT) enabled three-node differential decode-and-forward (DDF) relay networks, which adopt the power splitting (PS) protocol at the relay. The use of non-coherent differential modulation eliminates the need for sending training symbols to estimate the instantaneous channel state informations (CSIs) at all network nodes, and therefore improves the power efficiency, as compared with the coherent modulation. However, performance analysis results are not yet available for the state-of-the-art detectors such as the approximate maximum-likelihood detector. Existing works rely on Monte-Carlo simulation to show that there exists an optimal PS ratio that minimizes the overall SER. In this work, we propose a near-optimal detector with linear complexity with respect to the modulation size. We derive an accurate approximate SER expression, based on which the optimal PS ratio can be accurately estimated without requiring any Monte-Carlo simulation.
\end{abstract}

\section{Introduction}
The radio frequency (RF) signal has been widely used as the carrier for wireless information transmission (WIT). It has also become a new source for energy harvesting (EH) in the wireless power transfer (WPT) process \cite{ref0-zhou2013wireless}. In recent years, simultaneous wireless information and power transfer (SWIPT) has emerged as a promising technology to use the RF signal for WIT and WPT at the same time \cite{intro-tang2018energy}. SWIPT is an essential technology for various wireless systems to support different applications (see \cite{swipt-8114544} and references therein), for example, for 5G communications to support the Internet of Things (IoT) applications \cite{intro-IoT-1-yang2017impact}. Most of the devices deployed in the IoT networks are of small sizes and low-powered, and harvesting energy from the RF signals can be a sustainable solution to provide them with convenient energy supplies \cite{intro-IoT-2-perera2018simultaneous}. 

In the SWIPT-enabled relay networks, the relay plays both the roles of EH for WPT and information processing for WIT. Two main receiver architectures are available for practical use at the relay \cite{intro-zhang2013mimo}, namely, the power splitting (PS) and time switching (TS) architectures. For the PS protocol, the received signal is separated in two portions which are used for EH and information processing operations, respectively. For the TS protocol, these two operations are performed in a time-division fashion.

One of the key challenges in the SWIPT-enabled relay networks is that the RF-powered relay nodes are energy-constrained, which restraints the use of high power consumption coding and decoding, modulation and demodulation techniques. The existing works, such as \cite{ref0-zhou2013wireless} \cite{intro-boshkovska2017secure} \cite{intro-ref6-atapattu2016optimal} \cite{ref1-ye2018optimal}, use a coherent setup and assume that the instantaneous channel state informations (CSIs) are available at the receiving nodes. However, to acquire the instantaneous CSIs requires frequent channel estimations, which causes additional power consumption and is not friendly for such energy-constrained networks. Moreover, the additionally consumed power has negative impact on the future data relaying when the total power budget is fixed \cite{dm-df-ser-1-liu2015energy}. To address this issue, the power-efficient non-coherent differential modulation (DM) technique, which eliminates the channel estimation requirements, has become an attractive solution.  


Several works have been done to study the performance of SWIPT-enabled differential decode-and-forward (DDF) and amplify-and-forward (DAF) relay networks in the literature (see \cite{dm-af-1-mohjazi2018performance, dm-af-ber-lou2017exact,dm-af-ser-1-liu2015noncoherent,dm-df-ser-1-liu2015energy}). In the information-theoretic perspective, for DAF, several performance metrics were studied in \cite{dm-af-1-mohjazi2018performance}, such as the outage probability,
achievable throughput and average symbol error rate (SER). The performance of selection combining was studied in \cite{dm-af-ber-lou2017exact}. In the communication-theoretic perspective, the SER performances of the maximum likelihood detectors (MLDs) based on the PS and TS protocols were studied in \cite{dm-af-ser-1-liu2015noncoherent} and \cite{dm-df-ser-1-liu2015energy}, respectively for DAF and DDF relay networks, and the approximate MLDs with lower complexities were also obtained. These detectors serve as good performance benchmarks. However, their performance analysis results are not yet available in the literature, and the SERs of the respective systems were studied via Monte-Carlo simulation \cite{dm-af-1-mohjazi2018performance}. This is possibly due to the non-closed-form detection metrics of the detectors. For example, the metrics involve functions such as the modified Bessel function of the second kind, and even integral calculation.

To the best of our knowledge, for such SWIPT-enabled DDF relay networks, the SER analysis associated with a near-optimal detection scheme has not been studied in the literature. Motivated by this, we propose a near-optimal detector with a closed-form metric whose SER performance closely approaches that of the MLD in \cite{dm-df-ser-1-liu2015energy}, adopting the PS protocol at the relay. For the SER analysis, an accurate approximate SER expression is derived. In addition, we propose two methods to numerically estimate the optimal value of the PS ratio that minimizes the SER, and both methods are verified by simulations to be quite accurate. 

The rest of the paper is organized as follows. Section \ref{sec:sys-mod} presents the system model. The detector and approximate SER expression are proposed in Section \ref{sec:det-bound}. Section \ref{sec:ana} analyzes this expression and studies the optimal PS ratio. Section \ref{sec:simu} presents simulation
results followed by conclusions in Section \ref{sec:con}. Related proofs are provided in the Appendix.

\textit{Notation:} $p_X(X)$ denotes the probability density function of the random variable $X$, and $\Pr[X]$ is the probability of an event $X$. $\mathbb{E}[X]$ represents the expected value of $X$. $\Re \{x\}$ denotes the real part of a complex number $x$. $Q(x) \triangleq \frac{1}{\sqrt{2 \pi}} \int_{x}^{\infty} \exp (-z^2/2) dz$.
\section{System Model}
\label{sec:sys-mod}
\begin{figure}[t]
	\centering
	\includegraphics[width=3.5in]{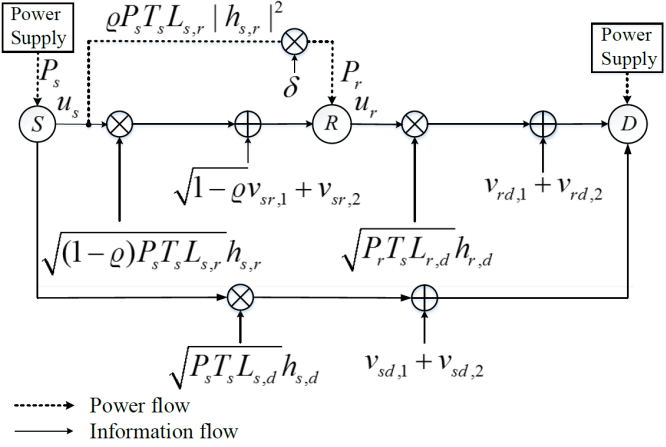}
	\caption{The system model of the $3$-node SWIPT-enabled PS-based DDF relay network, where the solid and dashed arrow lines denote the information and the power flows, respectively.}
	\label{fig:channel-model-PS}
\end{figure} %

We consider a $3$-node SWIPT-enabled PS-based DDF relay network with one source ($S$), one half-duplex relay ($R$) and one destination ($D$). The system model is shown in Fig.~\ref{fig:channel-model-PS}, where the information and power flows are shown using the solid and dashed arrow lines, respectively. Assume $R$ has no CSI, while $D$ has only the statistical CSI of the $S-R$ link but no CSI of the other links. $S$ and $D$ have dedicated energy sources such as a battery or power grid, and $S$ transmits its message with a constant power $P_s$. However, $R$ has no power supply and can only harvest energy from the received signals from $S$ for detection and transmission. For the $I-J$ link, $(I,J) \in \{(s,r), (s,d), (r,d)\}$, assume small-scale Rayleigh fading $h_{I,J}$ and large-scale path loss $L_{I,J}$. Let $v_{IJ,1} \sim \mathcal{CN}(0, N_{IJ,1})$ denote the complex additive
white Gaussian noise (AWGN) at the receive antenna and $v_{IJ,2} \sim \mathcal{CN}(0, N_{IJ,2})$ denote the complex AWGN due to the circuit.

One symbol transmission consists of two time slots each with duration $T_s = \frac{T}{2}$. In the first time slot, $S$ transmits its signal to $R$ and $D$. In the second time slot, $R$ transmits its detected symbol to $D$ while $S$ remains silent. Differential PSK (DPSK) is used. Let $S$ select a symbol $x_s$ from the $M$-PSK alphabet, defined as $\mathcal{X} \triangleq  \{x_m = e^{j2 \pi (m-1)/M}, m=1,2, \cdots ,M\}$, with equal probability. $x_s$ is differentially encoded as $u_s$ for transmission, and $R$ uses the same approach to encode $x_r$ as $u_r$ for transmission. For the $k$-th symbol, we can write the differential encoding operation as $u_I[k] =  u_I[k-1] x_I[k]$, $I \in \{s,r\}$, $k=1,2,3,\cdots$, with the initialization symbol defined as $u_I[0]=1$ (c.f. \cite{dm-df-ser-1-liu2015energy}). 

The PS protocol is used at $R$ for EH with the PS ratio $\varrho \in (0,1)$. Specifically, $R$ splits its received signal from $S$ into two portions, a $\varrho$ portion for EH and a $1-\varrho$ portion for information detection (ID). We use the linear EH model here for simplicity, and leave the study of the non-linear model (such as that described in \cite{eh-non-li-boshkovska2015practical}) as future work. The harvested power at $R$ is 
$
P_r = \frac{\delta \varrho P_s L_{s,r} |h_{s,r}|^2 T_s}{T_s} = \delta \varrho P_s L_{s,r} |h_{s,r}|^2
$ with $\delta \in (0,1]$ as the relay power conversion efficiency. The impact of the quality of the $S-R$ link on the transmission of the $R-D$ link is obvious since the relay transmission power $P_r$ is directly affected by the term $L_{s,r} |h_{s,r}|^2$, which is a measure of the $S-R$ link quality. Received signals at $R$ for ID and at $D$ are
\begin{align}
	y_{s,r}[k] = & \sqrt{(1-\varrho)P_s T_s L_{s,r}} h_{s,r} u_s[k] + \notag \\
	& \sqrt{1-\varrho}v_{sr,1}[k] + v_{sr,2}[k] ,\\
	y_{I,d}[k] = & \sqrt{P_I T_s L_{I,d}} h_{I,d} u_I[k] + \notag \\
	&  v_{Id,1}[k] + v_{Id,2}[k], I \in \{s,r\}. 
\end{align}

For DM, by assuming the channel coefficients remain unchanged for at least two consecutive symbol intervals, the received signals can also be written as
\begin{align} \label{eq:p2p}
	y_{I,J}[k] = & y_{I,J}[k-1] x_I[k] + n_{I,J},
\end{align}
where
$
n_{s,r} =  \sqrt{1-\varrho}v_{sr,1}[k] + v_{sr,2}[k] -  x_s[k](\sqrt{1-\varrho}v_{sr,1}[k-1] + v_{sr,2}[k-1]) \sim  \mathcal{CN} (0, 2(1-\varrho) N_{sr,1} + 2 N_{sr,2}) $, and $n_{I,d} =  v_{Id,1}[k] + v_{Id,2}[k]-  x_I[k](v_{Id,1}[k-1] + v_{Id,2}[k-1]) \sim  \mathcal{CN} (0, 2 N_{Id,1} + 2 N_{Id,2}) 
$. Note that for the detection at $R$, we adopt \cite[eq. (27)]{dm-df-ser-1-liu2015energy}, which is performed based on the relation of the two consecutively received symbols as shown in \eqref{eq:p2p}, and thus requires no CSI of the $S-R$ link.

\section{Proposed Detector and Approximate SER}
\label{sec:det-bound}
\subsection{Proposed Detector at the Destination}
The optimal MLD for DM should find the source symbol that maximizes the conditional joint probability density of the received signals as 
$\max_{x_s \in \mathcal{X}} f(y_{s,d}[k]|x_s, y_{s,d}[k-1]) \sum_{x_r \in \mathcal{X}} \Pr(x_r|x_s) f(y_{r,d}[k]|x_r,y_{r,d}[k-1])$.

By using the average SER (denoted as $\epsilon$) of ID at $R$ \cite[eq. (33)]{dm-df-ser-1-liu2015energy} to approximate the transition probability term $\Pr(x_r|x_s)$, and applying the widely-used max approximation, which gives the excellent performance, especially at high SNR (c.f. \cite{qian2017near}, \cite{kim2015low}, \cite{ours-icassp}), we obtain a near-optimal detection metric as 
\begin{align}
	\max_{x_s \in \mathcal{X}} &   \{ f(y_{s,d}[k]|x_s, y_{s,d}[k-1]) \max \{ 
	\notag \\
	& (1-\epsilon)  f(y_{r,d}[k]|x_s,y_{r,d}[k-1]) 
	, \notag \\
	& \epsilon/(M-1) \max_{ \substack{x_r \in \mathcal{X}, x_r \neq x_s} } f(y_{r,d}[k]|x_r,y_{r,d}[k-1])  \} \} . \notag
\end{align}

Further, based on the observation that $\epsilon < 0.5$ is sufficient to ensure $1-\epsilon > \epsilon/ (M-1)$, we remove the constraint $x_r \neq x_s$ and finally develop the proposed near-optimal detector, which is performed based on 
	\begin{align}		
	\hat{x}_s  
	= &  \arg  \min_{x_s \in \mathcal{X}}  \{
	\Re \{ y_{s,d}^*[k] y_{s,d}[k-1] x_s\} /(N_{sd,1}+ N_{sd,2}) +  \notag \\ & \min \{ 
	\Re \{ y_{r,d}^*[k] y_{r,d}[k-1] x_s\}/( N_{rd,1} + N_{rd,2})+\eta,  \notag \\
	&  \min_{x_r \in \mathcal{X}} \Re \{ y_{r,d}^*[k] y_{r,d}[k-1] x_r\}/( N_{rd,1} + N_{rd,2}) 
	\}
	\} ,  \notag
	\end{align}
where $\eta  \triangleq \log\frac{(1-\epsilon)(M-1)}{\epsilon} $. The complexity of this detector is linear with respect to the modulation size $M$, because the enumerations over $x_s$ and $x_r$ are decoupled. 

\subsection{Approximate SER for the Detector}
Consider the case  $N_{IJ,1}=N_{IJ,2}=N_0/2$ for all links hereafter for simplicity, and the instantaneous SNR of the $I-J$ link is defined as $\gamma_{I,J} \triangleq \frac{P_s |h_{I,J}|^2}{N_0}$. Denote the real Gaussian random variables associated with the detection metric of the $I-d$ link, $I \in \{s,r\}$, as
\begin{align}
\omega_{I,d} (z_1, z_2) 
= &  \Re \{ y_{I,d}^*[k] y_{I,d}[k-1] (z_2-z_1)\} /N_0 \notag \\
\sim & \mathcal{N} (u_{I,d}(z_1, z_2), W_{I,d} (z_1, z_2)), \label{eq:omega}
\end{align}
and we have
\begin{align}
u_{I,d}(z_1, z_2) = & T_s L_{I,d} \Re \{
x_{I}^*(z_2-z_1)
\} \frac{P_I |h_{I,d}|^2}{N_0} , \label{eq:w-mean}\\
W_{I,d}(z_1, z_2) \approx  & T_s L_{I,d} |z_2-z_1|^2 \frac{P_I |h_{I,d}|^2}{N_0}. \label{eq:w-var}
\end{align}
The mean in \eqref{eq:w-mean} can be obtained by substitutions. For the approximate variance in \eqref{eq:w-var}, the approximation is due to
\begin{align}
|y_{I,d}[k]|^2  = & |\sqrt{P_I T_s L_{I,d}} h_{I,d} u_I[k] + n_{I,d}[k] + v_{I,d}[k]|^2 \notag \\
\approx & P_I T_s L_{I,d} |h_{I,d}|^2 , \label{eq:approx}
\end{align}
where high order noise terms are ignored \cite{ref1-bhatnagar2012decode}. 

Using the defined variables in \eqref{eq:omega}, an approximate SER is derived in Appendix \ref{proof-prop-1}, and presented in Proposition \ref{prop: 1}, where $\check{\gamma} \triangleq [\gamma_{s,d}, \gamma_{r,d},\gamma_{s,r}]$, $g_{s,d} \triangleq  \sin^2  \left(\frac{\pi}{M}\right) T_s L_{s,d} $ and $ g_{r,d} \triangleq  \sin^2 \left(\frac{\pi}{M}\right)   T_s L_{s,r} L_{r,d}$. $\mathcal{P}_C (\check{\gamma})$ and $\mathcal{P}_E (\check{\gamma})$ characterize the conditional SER performances for the two scenarios where the relay detects correctly and wrongly, respectively.
\begin{proposition}\label{prop: 1}
	The overall SER of the proposed near-optimal detector is accurately approximately by $  \mathcal{P}_C (\check{\gamma}) + \mathcal{P}_E  (  \check{\gamma} )$ for $M>2$, and $ \frac{1}{2} \mathcal{P}_C (\check{\gamma}) + \frac{1}{2} \mathcal{P}_E  (  \check{\gamma} )$ for $M=2$, where
	\begin{align}
	\mathcal{P}_C(\check{\gamma})
	\triangleq & 2 (1-\epsilon) Q 
	\left( 
	\sqrt{g_{s,d} \gamma_{s,d} +  \varrho  \delta g_{r,d}|h_{s,r}|^2 \gamma_{r,d}}
	\right)   + \notag \\ &  2(1-\epsilon) Q \left(
	\sqrt{ g_{s,d} \gamma_{s,d}} + \frac{\eta}{2}
	\frac{1}{\sqrt{  g_{s,d} \gamma_{s,d} }}
	\right)  \label{eq:PC} 
	\end{align}	
	and
	\begin{align}
	\mathcal{P}_E(\check{\gamma}) 
	\triangleq &  
	\frac{2 \epsilon}{M-1}  Q \left(
	\sqrt{ g_{s,d} \gamma_{s,d}} - \frac{\eta}{2}
	\frac{1}{\sqrt{  g_{s,d} \gamma_{s,d} }}
	\right)  + \notag \\ & 2 \epsilon 
	Q \left(
	\sqrt{ g_{s,d} \gamma_{s,d}}
	\right)  . \label{eq:PE}	
	\end{align}
\end{proposition}
 
 Correspondingly, the overall average SER can be obtained by averaging over the channel gains as $P_E + P_C$ with $
 P_{t} =	\int \mathcal{P}_{t}(\check{\gamma} )
 p_{\check{\gamma}} (\check{\gamma}) d \check{\gamma}$, where $p_{\check{\gamma}}(\check{\gamma})$ denotes the joint probability density function of $\check{\gamma}$. It is verified in Section \ref{sec:simu} that the proposed approximate SER expression is quite accurate for not too low SNR values. Therefore, it is a good approximation for the actual SER of the system, and will be used for the SER analysis in the next section.
\section{SER Analysis}
\label{sec:ana}
\subsection{SER Performance Trade-off}
Useful insights can be drawn from the proposed approximate SER expression in various aspects. As an example, we present the trade-off between the conditional SERs of the two scenarios where the relay detects correctly and wrongly as a function of $\varrho$ in this subsection.

The EH relay system should take advantage of a $S-R$ link with good quality, and therefore here we assume sufficiently high average SNR of the $S-R$ link, then there is $\epsilon  \rightarrow 0 $ and $\eta \approx  \log \frac{1}{\epsilon} \rightarrow \infty $. In this case, $\mathcal{P}_C (\check{\gamma})$ and $\mathcal{P}_E (\check{\gamma})$ can be further approximated using the dominating terms $\tilde{\mathcal{P}}_C(\check{\gamma})$ and $\tilde{\mathcal{P}}_E(\check{\gamma})$, respectively, as 
\begin{align}
 \tilde{\mathcal{P}}_C (\check{\gamma}) 
\triangleq & 	2(1-\epsilon) Q 
\left( 
\sqrt{g_{s,d} \gamma_{s,d} +  \varrho  \delta g_{r,d}|h_{s,r}|^2 \gamma_{r,d}}
\right) , \\
\tilde{\mathcal{P}}_E  (  \check{\gamma} ) 
\triangleq &    
\frac{2\epsilon}{M-1}  Q \left(
\sqrt{ g_{s,d} \gamma_{s,d}} - \frac{\eta}{2}
\frac{1}{\sqrt{  g_{s,d} \gamma_{s,d} }}
\right)  .
\end{align}

We want to emphasize that both $\epsilon$ and $\eta$ are functions of $\varrho$, and therefore $\tilde{\mathcal{P}}_C (\check{\gamma})$ and $\tilde{\mathcal{P}}_E (\check{\gamma})$ are functions of $\varrho$. By proving in Appendix \ref{appen-mono} that $\tilde{\mathcal{P}}_C( \check{\gamma} ) $ and $ \tilde{\mathcal{P}}_E(  \check{\gamma} ) $ are monotonically decreasing and increasing in $\varrho$, respectively, we can see that there exists a trade-off between the conditional SERs of the aforementioned two scenarios. One possible explanation is that increasing the PS ratio $\varrho$ will increase the relay transmission power $P_r$. For scenario one where the detection at $R$ is correct, the reliability of the overall $S-R-D$ link is increased, and the error probability decreases, while for scenario two where the detection at $R$ is wrong, increasing $P_r$ encourages error propagation, which in turn undermines the network reliability and increases the error probability. The results suggest that a good trade-off can potentially be achieved by adjusting $\varrho$. 
\subsection{Optimized Power Splitting Ratio}
Simulation results in \cite{dm-df-ser-1-liu2015energy} show that there is an optimal value of $\varrho$ that minimizes the SER. However, the task of finding the closed-form expression for this optimal value appears intractable. To address this issue, in this subsection, we propose two methods to estimate this optimal PS ratio off-line numerically. 


%

One method is by calculating the minimum of the average approximate SER, i.e., $P_E + P_C$, using software packages such as cvx with MATLAB; see \cite{cvx-grant2014cvx} for details. However, double integral calculations are required. To save the computational effort, we propose a second method by equivalently studying the zero of its derivative with respect to $\varrho$, i.e., the zero of $\frac{\partial P_C}{\partial \varrho} + \frac{\partial P_E}{\partial \varrho}$. 

 Approximating $Q(z) \approx \frac{1}{2}e^{-\frac{z^2}{2}}, z >0$, and after some mathematical manipulations, we have for the average approximate SER that $P_C \approx  (1-\epsilon)(Z_1 + Z_2)$ and $P_E \approx  \frac{\epsilon  Z_3}{M-1}  +  \frac{\epsilon }{ g_{s,d} \bar{\gamma}_{s,d}+2}$
 with $Z_1 =  a_1 \sqrt{2\eta} \exp (-2b_1 \eta) $, $Z_2 =    \frac{a_2}{\varrho} \ln \left(
 1+b_2 \varrho \right)$ and $Z_3 =  \exp (\eta) Z_1$. Then based on the chain rule of derivative, we have
 \begin{small}
 	\begin{align}
 	\frac{\partial P_C}{\partial \varrho} \approx & - \frac{\partial \epsilon}{\partial \varrho} (Z_1 + Z_2) + (1-\epsilon) 
 	\left(
 	\frac{\partial Z_1 }{\partial \eta } \frac{\partial \eta }{\partial \epsilon} \frac{\partial \epsilon}{\partial \varrho} + 
 	\frac{\partial Z_2 }{\partial \varrho }
 	\right), \notag  \\
 	\frac{\partial P_E}{\partial \varrho} \approx & \frac{Z_3}{M-1} \frac{\partial \epsilon}{\partial \varrho}  + \frac{\epsilon}{M-1}   \frac{\partial Z_3 }{\partial \eta } \frac{\partial \eta }{\partial \epsilon} \frac{\partial \epsilon}{\partial \varrho} + \frac{1 }{ g_{s,d} \bar{\gamma}_{s,d}+2} \frac{\partial \epsilon}{\partial \varrho} \notag \\
 	= & \left( \frac{\epsilon }{M-1}  \frac{\partial Z_3 }{\partial \eta } \frac{\partial \eta }{\partial \epsilon}  + \frac{ Z_3}{M-1}  + \frac{1 }{ g_{s,d} \bar{\gamma}_{s,d}+2} \right)  \frac{\partial \epsilon}{\partial \rho(  \varrho )}
 	\frac{\partial \rho(  \varrho ) }{\partial \varrho}, \notag
 	\end{align}
 \end{small}where the function $\rho(  \varrho )$ is defined as $\rho(  \varrho ) \triangleq \frac{1-\varrho}{ 2-\varrho }$, and some auxiliary variables not related to $\varrho$ are defined as $a_1 = \frac{\sqrt{\pi}(2 g_{s,d})^{-\frac{1}{4}}}{4 \bar{\gamma}_{s,d}} \left( g_{s,d}/2+\bar{\gamma}_{s,d}^{-1} \right)^{-\frac{3}{4}}$, $a_2 = \frac{ 2 \frac{P_s}{N_0}}{\delta g_{r,d} ( g_{s,d} \bar{\gamma}_{s,d}+2) \bar{\gamma}_{s,r} \bar{\gamma}_{r,d}}$, $b_1 = \frac{1}{4} + \frac{ \sqrt{ g_{s,d}/2+\bar{\gamma}_{s,d}^{-1}}}{2  \sqrt{2 g_{s,d}}}$ and $b_2 = \frac{ \delta g_{r,d}  \bar{\gamma}_{s,r} \bar{\gamma}_{r,d}}{2 \frac{P_s}{N_0}}$. By further substitutions, we have that the derivative, i.e., $\frac{\partial P_C}{\partial \varrho} + \frac{\partial P_E}{\partial \varrho}$, can be accurately approximated in closed-form using
\begin{align}
\frac{\partial P_C}{\partial \varrho} \approx &  \left( a_1 \sqrt{2\eta} \exp (-2b_1 \eta) + \frac{a_2}{\varrho} \ln \left(
1+b_2 \varrho \right)
\right)  \notag \\ & 
\sqrt{  \frac{ 1 + \rho(  \varrho )b_3 }{\rho(  \varrho ) b_3}  } \frac{ -a_3b_3 }{ 2 (1+ \rho(  \varrho ) b_3)^2(2-\varrho)^2}  + \notag \\
& 	\frac{ a_1}{ \epsilon }  \exp (-2b_1 \eta)
\left( -   b_1 \sqrt{2\eta}  +  \frac{1}{2\sqrt{2\eta}}   	\right)
 \notag \\ &
\sqrt{  \frac{ 1 + \rho(  \varrho )b_3 }{\rho(  \varrho ) b_3}  }  \frac{ -a_3b_3 }{ (1+ \rho(  \varrho ) b_3)^2(2-\varrho)^2}  + 
 \notag \\
& (1-\epsilon)
\left(
-\frac{a_2}{\varrho^2} \ln \left(
1+b_2 \varrho \right) + \frac{a_2 b_2}{\varrho(1+b_2 \varrho)} 
\right)  \label{eq:deriv-1},	\\
\frac{\partial P_E}{\partial \varrho} \approx &  \left[ \frac{1 }{ g_{s,d} \bar{\gamma}_{s,d}+2} + \frac{a_1}{M-1}   \exp (\eta) \exp (-2b_1 \eta) \right. \notag \\ & \left. \left(
\left( \frac{ \sqrt{2\eta} }{ 2 } - b_1 \sqrt{2\eta} + \frac{1}{ 2 \sqrt{2\eta} } 
\right)  \frac{-2}{  (1-\epsilon)} + \sqrt{2\eta}
\right)
\right] \notag \\
&  \sqrt{  \frac{ 1 + \rho(  \varrho )b_3 }{\rho(  \varrho ) b_3}  } \frac{ a_3b_3 }{ 2 (1+ \rho(  \varrho ) b_3)^2 (2-\varrho)^2}  \label{eq:deriv-2},
\end{align}
where $a_3 = 1.03 \sqrt{\frac{1+\cos \frac{\pi}{M}}{2 \cos \frac{\pi}{M}}} $, $a_4 = \frac{1 }{ g_{s,d} \bar{\gamma}_{s,d}+2}$ and $b_3 = (1-\cos \frac{\pi}{M}) T L_{s,r} \bar{\gamma}_{s,r}$.

Numerical results in Section \ref{sec:simu} show that the approximate derivative is monotonically increasing with $\varrho$ and has a unique zero, which well matches the minimum of the simulated SER.

\section{Numerical and Simulation Results}
\label{sec:simu}
\begin{figure}[t]
	\centering
	\includegraphics[width=0.48\textwidth]{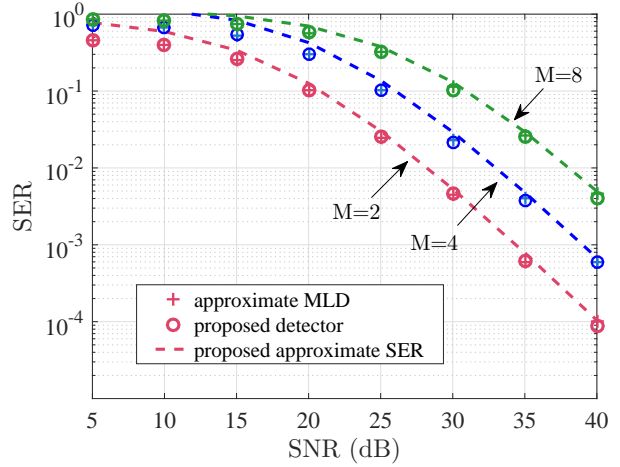}
	\caption{Comparison of different detectors and the proposed approximate SER expression with respect to the SNR (dB) when $\varrho=0.8$, for the 3-node SWIPT-enabled DDF relay networks with $M$-DPSK.}
	\label{fig:SER-2}
\end{figure}
In this section, numerical and simulation results are presented for evaluation. The near-optimal SER performance of the proposed detector and the accuracy of the approximate SER are shown. The accuracies of the proposed two ways for estimating the optimal PS ratio are also validated. The parameters are set based on \cite{dm-df-ser-1-liu2015energy}. The EH efficiency is set to $\delta=0.6$ without otherwise stated, and we use the bounded path-loss model as $L_{I,J}=\frac{1}{1+d_{I,J}^{2.7}}$ to ensure that the path-loss is strictly larger than one. The distances between the nodes are set as $d_{s,d}=3$, and $d_{s,r}=d_{r,d}=1.5$. The SER performance is parameterized by the transmitter SNR defined as SNR $\triangleq P_s/N_0$. The transmission rate is set as $R=\log_2 M$ for simplicity, which only depends on the modulation size $M$.

\begin{figure}[t] 
	\centering
	\subfigure[$M=2$, SNR = $30$ dB ]{
		\label{Fig.sub.1}
		\includegraphics[width=0.48\textwidth]{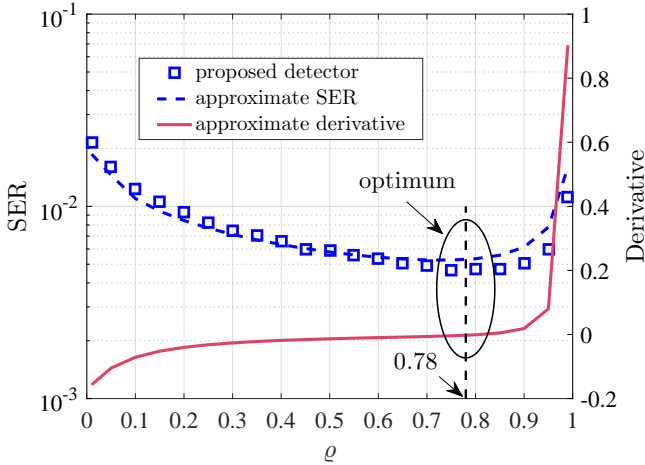}}
	\subfigure[$M=8$, SNR = $40$ dB]{
		\label{Fig.sub.2}
		\includegraphics[width=0.48\textwidth]{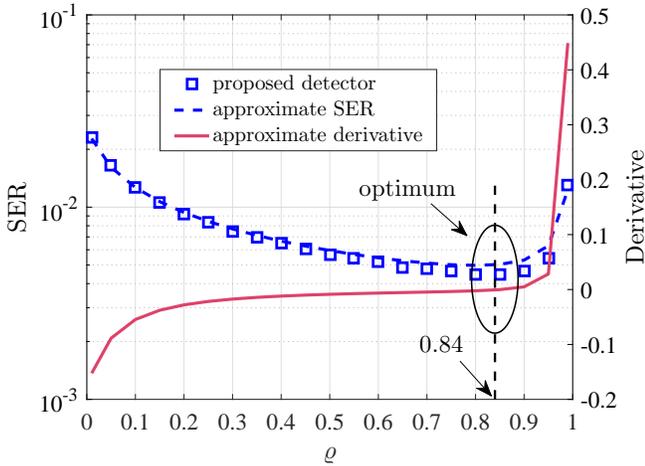}}
	\caption{The simulated SER of the proposed detector and the proposed approximate SER and derivative with respect to the PS ratio $\varrho$, for the 3-node SWIPT-enabled DDF relay networks with $M$-DPSK.}
	\label{fig:SER-1}
\end{figure}
Fig.~\ref{fig:SER-2} compares the SER performances of our detector and the state-of-the-art approximate MLD \cite{dm-df-ser-1-liu2015energy} for different transmission rates (modulation sizes), when $\varrho=0.8$ with respect to the SNR (dB). The proposed approximate SER is also simulated to show its accuracy. It is seen that the two detectors show similar SER performance, and both achieve the full diversity order of $2$. Because it has been verified in \cite{dm-df-ser-1-liu2015energy} that this approximate MLD is an excellent approximation for the exact MLD, the results verify that our detector is near-optimal. It is also notable that the approximate SER is quite accurate for not too low SNR values.	

Fig.~\ref{fig:SER-1} shows the simulated SER of the proposed detector, approximate SER and derivative, for DBPSK with $R=1$ (bps) at SNR = $30$ dB and $8$-DPSK with $R=3$ (bps) at SNR = $40$ dB. To show the SER and the derivative simultaneously, double y-axes is used with the SER value on the left y-axis and the derivative value on the right. There are several observations that can be made from Fig.~\ref{fig:SER-1}. The first is that the simulated SER has a unique minimum. The second is that the proposed approximate SER is quite accurate for all $\varrho \in (0,1)$ considered, and also shows a unique minimum. The third is that the derivative is monotonically increasing from negative to positive with $\varrho$, which suggests that the proposed approximate SER is convex in $\varrho$. Most notably, it is seen that the minimums of the simulated and approximate SERs and the zero of the approximate derivative are approximately the same (up to the second decimal digit). The optimal values are approximately $0.78$ and $0.84$, respectively, for $M=2$ and $8$. Therefore, both the approximate SER and derivative can be used to estimate the optimal PS ratio accurately. 

\begin{figure}[t]
	\centering
	\includegraphics[width=0.495\textwidth]{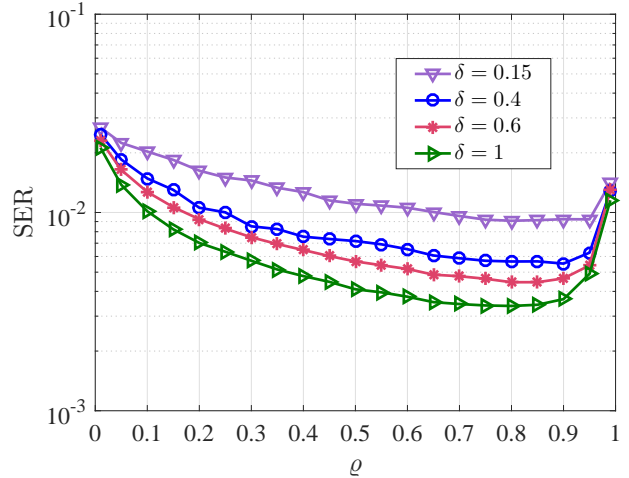}
	\caption{Comparison of the SER for different EH efficiency $\delta$ with respect to the PS ratio $\varrho$, using the proposed detector with $8$-DPSK at SNR $= 40$ dB.}
	\label{fig:SER-effi}
\end{figure}

Fig.~\ref{fig:SER-effi} compares the simulated SER for different values of the EH efficiency $\delta=0.15, 0.4, 0.6$ and $1$ using our detector. The modulation is $8$-DPSK and SNR $= 40$ dB. It can be observed that for a fixed $\varrho \in (0,1)$, the SER decreases with $\delta$. This is possibly because as $\delta$ increases, $R$ is capable of harvesting more power from the same received signals, and therefore the overall SER performance is improved. It is also notable that increasing the EH efficiency $\delta$ will shift the optimal value of $\varrho$ to left. An interpretation of this is that as $\delta$ increases, $R$ becomes more energy efficient, and therefore a smaller value of $\varrho$ is needed to maintain the same reliability as that of the previous. 
\section{Conclusion}
\label{sec:con}
In this paper, we have proposed a near-optimal detector with linear complexity with respect to the modulation size, and developed new SER performance results for the SWIPT-enabled PS-based DDF relay network. The state-of-the-art detectors are the MLD and the approximate MLD derived in \cite{dm-df-ser-1-liu2015energy}. They serve as good performance benchmarks. However, 
their performance analysis results are not available in the literature, possibly due to the complicated non-closed-form detection metrics involving functions such as the modified Bessel function. 
Our proposed detector has a closed-form metric, and its SER performance has been compared favorably with the approximate MLD. We have also proposed an approximate SER expression for our detector, and this expression has been shown to be rather accurate for all values of the PS ratio $\varrho$ considered, for not too small SNR values. Through analyzing this expression, we have presented the trade-off between the conditional SERs of the two scenarios where the relay detects correctly and wrongly as a function of $\varrho$. The results suggest that a good trade-off can potentially be achieved by adjusting $\varrho$. Moreover, we have proposed two methods for accurately estimating the optimal PS ratio that minimizes the SER. One is by finding the minimum of the explicit-form average approximate SER expression, which is straightforward but double integral calculation is need and is computationally expensive. The other is through finding the zero of the derived closed-form approximate derivative of the average approximate SER. Both methods have been verified to be quite accurate by simulations. 

\appendix
\subsection{Proof of Proposition \ref{prop: 1}}
\label{proof-prop-1}
Without loss of generality, assume $x_1$ is the source symbol and is wrongly detected to $x_v$ at $D$. Two competing symbol pairs are denoted as $(x_1, x_r)$ and $(x_v, x_u), x_v \neq x_1$.
	
For the scenario where the relay detects correctly, we have $x_r = x_1$. The problem of obtaining the dominating PEP terms can be formulated as
\begin{align} \label{eq:relay-correct}
\underset{x_v, x_u}{\max}  \quad & 
\left\{  \Pr [\omega_{s,d} (x_1, x_v) + \omega_{r,d} (x_1, x_v)>0 ] , \right. \notag \\
& \left.  \Pr [ \omega_{s,d} (x_1, x_v) + \omega_{r,d} (x_1, x_u) > \eta]
\right\} \notag \\
\subto \quad & (x_v,x_u) \in \mathcal{X}^2, x_v \neq x_1, x_u \neq x_v.
\end{align}
Our approach is to take all possible solutions to \eqref{eq:relay-correct} to formulate an approximate conditional SER expression $\mathcal{P}_C (\check{\gamma}) $. 

To maximize the first term in the objective, based on \eqref{eq:w-mean} and \eqref{eq:w-var}, we should maximize
\begin{align}
& \Pr [\omega_{s,d} (x_1, x_v) + \omega_{r,d} (x_1, x_v) >0 ] \notag \\
\approx  &  Q \left( \sqrt{\frac{-u_{s,d}(x_1, x_v)-u_{r,d}(x_1, x_v)}{2}}
\right) , \label{eq:case1}
\end{align}
and equivalently we should  $\minimize \quad -u_{s,d}(x_1, x_v)-u_{r,d}(x_1, x_v)$. Based on some calculations, we can obtain two possible solutions as $x_v \in \{x_2, x_M\}$. Similarly to maximize the second term, the problem is re-formulated as 
\begin{align}
\underset{x_v,x_u}{\min}	\quad  & \sqrt{-u_{s,d}(x_1, x_v)-u_{r,d}(x_1, x_u)}  + \notag \\ & \frac{\eta}{\sqrt{-u_{s,d}(x_1, x_v)-u_{r,d}(x_1, x_u)} } \notag \\
\subto \quad & (x_v,x_u) \in \mathcal{X}^2, x_v \neq x_1, x_u \neq x_v.
\end{align}

We make the assumption that $x_1$ is wrongly detected to its nearest neighbors at $D$ in this case, which is well justified when the relay detects correctly. After some calculations, the solution set is obtained as $
x_v \in \{x_2, x_M\}, x_u=x_1
$. Finally, $ \mathcal{P}_C (\check{\gamma})$ can be obtained using all dominating PEP terms. 

For the scenario where the relay detects wrongly, similarly to the previous case, we take all possible solutions to \eqref{eq:relay-wrong-1} and \eqref{eq:relay-wrong-2}, respectively, to formulate an approximate conditional SER expression $\mathcal{P}_E (\check{\gamma}) $.
\begin{align}
\underset{x_v,x_u}{\max} 
\quad	& \Pr [\omega_{s,d} (x_1, x_v) + \omega_{r,d}(x_r, x_u) > 0 ]  \notag \\
\subto \quad & (x_v,x_u) \in \mathcal{X}^2, x_v \neq x_1, x_u \neq x_v.
\label{eq:relay-wrong-1} \\	
\underset{x_v}{\max}  \quad	& \Pr [\omega_{s,d} (x_1, x_v) + \omega_{r,d}(x_r, x_v) > -\eta ] \notag \\
\subto \quad & x_v \in \mathcal{X}, x_v \neq x_1.
\label{eq:relay-wrong-2}
\end{align}
After some calculations, the possible solutions are
obtained as $
x_v \in \{x_2, x_M\}, x_u=x_r, x_u \neq x_v
$, and $x_v=x_r, x_v \in \{x_2, x_M\}$, for \eqref{eq:relay-wrong-1} and \eqref{eq:relay-wrong-2}, respectively. $\mathcal{P}_E  (  \check{\gamma} )$ can be obtained accordingly.
\subsection{Proof of Monotonicity of $\tilde{\mathcal{P}}_C( \check{\gamma} )$ and $\tilde{\mathcal{P}}_E( \check{\gamma} )$}
\label{appen-mono}
Based on the expressions of $\epsilon$ and $\eta$, there is $\frac{ \partial \epsilon}{  \partial \varrho } > 0$ and $\frac{ \partial \eta}{  \partial \varrho } < 0$. For $\tilde{\mathcal{P}}_C( \check{\gamma} )$, since both $1-\epsilon$ and $Q 
\left( 
\sqrt{g_{s,d} \gamma_{s,d} +  \varrho  \delta g_{r,d}|h_{s,r}|^2 \gamma_{r,d}}
\right)$ are positive and monotonically decreasing in $\varrho$, $\tilde{\mathcal{P}}_C( \check{\gamma} )$ is monotonically decreasing in $\varrho$. By taking the derivative of $\tilde{\mathcal{P}}_E( \check{\gamma} )$ with respect to $\varrho$, we have
\begin{align}
\frac{\partial \frac{\tilde{\mathcal{P}}_E( \check{\gamma} )}{\frac{2}{M-1}}}{\partial \varrho} 
= & \frac{-\exp \left(-
\frac{z_0^2}{2}
\right) }{ 2 (1-\epsilon) \sqrt{2\pi g_{s,d} \gamma_{s,d}} } \frac{\partial \epsilon }{\partial \varrho} +  Q 	\left( z_0
\right) \frac{\partial \epsilon }{\partial \varrho}, \label{eq:PEa-deriv}
\end{align}
where $ z_0 = \sqrt{ g_{s,d} \gamma_{s,d}} - \frac{\eta }{2   \sqrt{ g_{s,d} \gamma_{s,d}} } \overset{\eta \rightarrow \infty }{\approx} - \frac{\eta }{2   \sqrt{ g_{s,d} \gamma_{s,d}} } < 0$ . 

An accurate approximation as $Q(z_0) \approx \frac{1}{12} \exp \left(-\frac{z_0^2}{2}\right) + \frac{1}{4} \exp \left(-\frac{2}{3}z_0^2 \right), z_0 >0$ is applied to \eqref{eq:PEa-deriv}, then to prove $\frac{\partial \tilde{\mathcal{P}}_E( \check{\gamma} )}{\partial \varrho} > 0$ is equivalent to prove
\begin{align}
\frac{\frac{1}{ 2 \sqrt{ 2 \pi g_{s,d} \gamma_{s,d}} } \exp \left(-
	\frac{z_0^2}{2}
	\right) }{1-\frac{1}{12} \exp \left(-\frac{z_0^2}{2}\right) - \frac{1}{4} \exp \left(-\frac{2}{3}z_0^2 \right)} < 1-\epsilon \label{eq:mono-PE} .
\end{align}
For the left side of \eqref{eq:mono-PE}, when $\eta \rightarrow \infty $, it can be approximated as $\frac{1}{ 2 \sqrt{2 \pi g_{s,d} \gamma_{s,d}} } \exp \left(-
\frac{z_0^2}{2}
\right)$, of which the value approaches $0$,
while the value of the right side of \eqref{eq:mono-PE} approaches $1$. Therefore \eqref{eq:mono-PE} holds.
\bibliographystyle{IEEEtran} 
\bibliography{ref_swipt_dm}
\end{document}